\documentclass[prx,superscriptaddress,amsmath,amssymb,reprint]{revtex4-2}

\usepackage{braket}
\usepackage{comment}
\usepackage{graphicx}
\usepackage{float}
\usepackage{tikz}
\usepackage{dcolumn}
\usepackage{bm}
\usepackage{hyperref}
\usepackage{mathtools}
\usepackage [english]{babel}
\usepackage [autostyle, english = american]{csquotes}
\usepackage{comment}
\MakeOuterQuote{"}

\usepackage{graphicx}
\usepackage{amssymb}
\usepackage{amsbsy}
\usepackage{braket}
\usepackage{topcapt}
\usepackage{epstopdf}
\usepackage{bbm}
\usepackage{dirtytalk}
\usepackage[dvipsnames]{xcolor}
\usepackage{lipsum}
\usepackage{mathtools}
\usepackage{physics}
\newcommand{\la}{\langle}
\newcommand{\ra}{\rangle}

\newcommand{\be}{\begin{eqnarray}}
\newcommand{\ee}{\end{eqnarray}}
\newcommand{\cof}{\cos(\phi)}
\newcommand{\sif}{\sin(\phi)}
\newcommand{\cofs}{\cos^2(\phi)}
\newcommand{\sifs}{\sin^2(\phi)}
\newcommand{\sq}{\frac1{\sqrt2}}
\usepackage{mathtools}
\usepackage [english]{babel}
\usepackage [autostyle, english = american]{csquotes}

\begin{document}

\title{Evidence of Uncollapsed Quantum Amplitudes After Consecutive Measurements}

\author{Christoph Adami}
\email{adami@msu.edu}
\affiliation{Department of Physics and Astronomy, Michigan State University, East Lansing, MI 48824,USA}

\author{Lambert Giner}
\affiliation{Département de Physique et d’Astronomie, Université de Moncton, Moncton, New Brunswick E1A 3E9, Canada}

\author{Jeff S. Lundeen}
\affiliation{Department of Physics and Nexus for Quantum Technologies, University of Ottawa, Ottawa, Ontario K1N 6N5, Canada}

\author{Raphael A. Abrahao}
\email{raasps@rit.edu}
\affiliation{Department of Physics, University of Ottawa, Ottawa, Ontario K1N 6N5, Canada}
\affiliation{School of Physics and Astronomy, Center for Detectors, Future Photon Initiative, Rochester Institute of Technology, Rochester, New York 14623, USA}

\date{\today}                                           

\begin{abstract}
Two of the most common interpretations of quantum measurement disagree about the fate of quantum amplitudes after measurement, yet this disagreement has not previously led to experimentally distinguishable predictions. In the standard collapse picture, commonly linked to the Copenhagen interpretation of quantum mechanics, measurements eliminate unrealized amplitudes without leaving a memory. In contrast, in the unitary theory, the measurement device registers one of the possible outcomes while remaining part of an entangled state that continues to harbor the unrealized amplitudes. This persistence arises naturally under unitary evolution, since a measurement device that is part of an entangled system cannot serve as a faithful probe of the joint quantum state. Using single-photon measurements of a tunable quantum state, we experimentally show that these two theories make different predictions when three or more consecutive measurements are performed on the same quantum system. Analysis of the joint density matrix of the three measurements reveals coherence among them and supports the unitary theory of quantum measurement. When decoherence is explicitly introduced, the joint density matrix of the quantum system of interest and the apparatus becomes consistent with what a collapse theory would predict. This work clarifies the dynamics of consecutive quantum measurements and offers new insights into the interpretation of quantum measurements.
\end{abstract}

\maketitle
\section{Introduction}

What happens to the probability amplitudes of a quantum wavefunction after a measurement is still one of the great mysteries of quantum physics. From an experimental point of view, a projective measurement performed on a quantum superposition of states invariably places the measurement device into one of the measurement operator's eigenstates. According to the Copenhagen interpretation, observing the measurement device in one of the measurement operator's eigenstates allows us to infer that the quantum state {\em itself} has collapsed from the initial superposition into the same measurement device eigenstate. In other words, it is assumed that after the measurement, the quantum system and the readout device will be perfectly correlated so that the readout is a diagnostic of the quantum state. While for classical measurements we can be assured that the measurement device reflects the state of the measured system, the same cannot be said for quantum systems on account of the quantum no-cloning theorem~\cite{WoottersZurek1982,Dieks1982,NielsenChuang_Book}.

 What then is the nature of the quantum superposition after measurement? According to the relative-state interpretation~\cite{vonNeumann1932,Everett1957}, quantum superpositions are maintained, implying that the ``unrealized'' amplitudes of the original wavefunction (those incompatible with the state of the measurement device) still exist after measurement, but are not registered within the device because a classical device can only be found in one eigenstate at the time (see note~\footnote{Common interpretations of Everett's theory assert that at each measurement event the universe is branching into as many copies as there are unrealized amplitudes, the so-called ``Many-Worlds'' interpretation of quantum mechanics. Such an interpretation, however, is unnecessary once it is understood that the state of the device is not a diagnostic of the quantum state. As a consequence, the measurement device can point to an eigenstate while the wavefunction continues to be in a 
  superposition~\protect\cite{GlickAdami2020}, and no branching into multiverses needs to be assumed.}). 
 Thus, we should not assume that the state of the measurement device is a diagnostic of the quantum state of interest, a 1-to-1 map. 

The two most common interpretations of the quantum measurement process cannot both be correct descriptions of reality, because the underlying quantum dynamics are fundamentally different. In a collapse picture of measurement, an unknown non-unitary process destroys unrealized amplitudes. In contrast, in the relative-state formulation, these amplitudes continue to be present even though there is no trace of all but one of them in our measurement device: the one corresponding to the measured outcome. 

If both interpretations lead to exactly the same observable predictions, then perhaps the purported differences in the underlying description of reality could be considered a matter of preference. Instead, here we show that these differences are measurable as long as more than two consecutive measurements are carried out on the same quantum system, and any quantum coherence between measurements is maintained. We experimentally demonstrate that one of the foundational assumptions, namely that a projective quantum measurement will invariably put the quantum state into one of its eigenstates, is at odds with our experimental evidence, and the predictions of the standard collapse postulate are only verified in the case that decoherence is allowed.

Below, we will refer to the relative-state formulation of quantum measurement as the {\em unitary theory} of measurement (see note~\footnote{It is worth pointing out that Everett's relative state picture of quantum measurement, which is different than the Many-Worlds interpretation of Everett and DeWitt, is essentially von Neumann's model from 1932~\protect\cite{vonNeumann1932}, but without the ``second stage'' of measurement that forces the quantum system into one of the eigenstates of the quantum system.}), while we use \textit{collapse theory} to refer to the standard Copenhagen view of the collapse of the wavefunction (even though the idea of wavefunction reduction is usually attributed to Heisenberg~\cite{Heisenberg1927}).

In the following, we briefly summarize the application of the unitary theory to consecutive measurements of the same quantum system~\cite{GlickAdami2020}, in order to highlight the differences in predictions between the collapse and the unitary theories for three consecutive measurements. We then describe an experiment designed to probe this difference and present results that support the unitary theory. We close by discussing our results and some foundational aspects of quantum mechanics.

\section{Unitary theory of consecutive measurements} \label{sec:theory}
\begin{figure}[] 
   \centering
   \includegraphics[width=1.0\columnwidth]{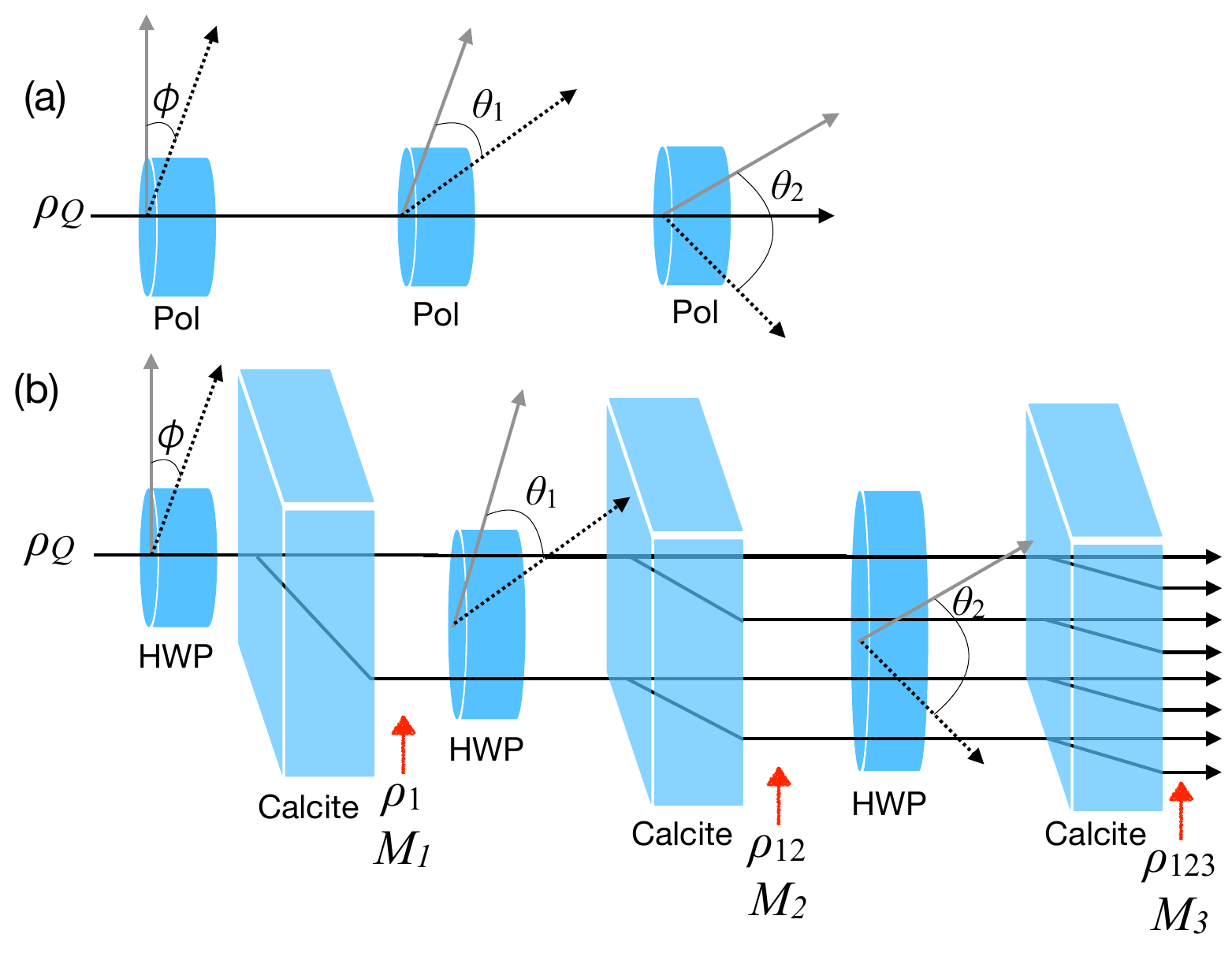} 
   \caption{(a) Consecutive measurements of a quantum state $\rho_Q$, showing rotation of the measurement basis by angles $\theta_1$ and $\theta_2$ using polarizers. (b) Experimental realization of the scheme using half-waveplate (HWP), calcite crystals and followed by photon detection. $\rho_Q$ denotes the initial quantum state, $\rho_1$ the density matrix of the first readout $M_1$, $\rho_{12}$ the joint state of the first two readouts after $M_2$, and $\rho_{123}$ the combined state of all three readouts following $M_3$.  
    }
   \label{measdevs}
\end{figure}

We describe consecutive measurements of a single qubit prepared in an arbitrary state $\rho_Q$, ranging from pure to completely mixed. The consecutive measurements are performed with three auxiliary readouts (pointers) such that the measurement bases are oriented at given angles with respect to each other (Fig.~\ref{measdevs}). A general description of consecutive measurements of qudits prepared in arbitrary states in the unitary formalism can be found in Ref.~\cite{GlickAdami2020}.

In our implementation, the quantum system of interest is encoded in the polarization degree of freedom, while the readout corresponds to the spatial path of the photon. Measuring the initial polarization state with an auxiliary readout system $M_1$ produces the density matrix
\begin{equation}
\rho_1 = \begin{pmatrix}
\cos^2(\phi) & 0 \\
0 & \sin^2(\phi)
\end{pmatrix},
\label{rho1}
\end{equation}
which represents the reduced density matrix of the path degree of freedom in the measurement eigenbasis, obtained after tracing out the polarization of the quantum system. The angle $\phi$ controls the initial mixedness of the polarization state, ranging from a pure state ($\phi = 0$) to a maximally mixed state ($\phi = \pi/4$).

If this measurement is performed non-destructively, not to be confused with a quantum non-demolition (QND) measurement~\cite{haroche2006book}, we can ask what happens when the same quantum state is measured again in a basis rotated by a finite angle relative to the first. In our setup, such strong consecutive measurements can be realized using optical Fourier transforms applied to a path-encoded quantum state~\cite{Curic19}. This sequence of measurements is directly analogous to sequential Stern–Gerlach experiments, as discussed by Sakurai and Napolitano~\cite{sakuraimodernQM}, Feynman~\cite{feynmanlectures}, Townsend~\cite{townsend2000modern}, and McIntyre~\cite{mcintyre2022quantum}. 

One of the most pronounced differences between the collapse predictions and the unitary theory predictions occurs when measurements are performed at orthogonal angles with respect to each other (a derivation of the density matrices in the unitary formalism using arbitrary angles can be found in Ref.~\cite{GlickAdami2020}).

A second measurement by a readout system $M_2$, oriented at $\theta_1=\pi/2$ relative to $M_1$, yields the joint density matrix for the quantum system of interest and the readout in the measurement basis:
\begin{eqnarray}
    \rho_{12}&=&    \frac12
    \begin{pmatrix}
    \cos^2(\phi) & 0 & 0 & 0 \\
      0 & \cos^2(\phi) & 0 & 0 \\
      0 & 0 & \sin^2(\phi) & 0\\
      0 & 0 & 0 & \sin^2(\phi)
    \end{pmatrix}\nonumber \\
    &=& \begin{pmatrix}
      \cos^2(\phi) & 0 \\
      0 & \sin^2(\phi) \\
   \end{pmatrix}\otimes  \begin{pmatrix}
      \frac12 & 0 \\
      0 & \frac12 \\
   \end{pmatrix}. \label{rho_12}
\end{eqnarray}
This density matrix is diagonal in the measurement basis, indicating that the joint state of the two readouts $M_1M_2$ is incoherent, just as one would expect for a pair of classical macroscopic objects~\cite{GlickAdami2020}.
If instead we trace the joint state over the two measurements, we obtain a reduced density matrix $\rho_Q$ aligned with the eigenstates of $M_2$, suggesting that the quantum state’s wavefunction has been reduced and that a wavefunction collapse occurred.

The density matrix in Eq.~(\ref{rho_12}) reproduces the predictions common to all existing interpretations of quantum measurement. For a fully mixed initial state ($\phi = \pi/4$), all four outcomes of the two binary measurements occur with equal probability $1/4$ under both collapse and unitary descriptions. However, because the unitary formalism does not invoke a wavefunction reduction, performing a {\em third} measurement on the same quantum state at a different orientation can reveal whether such a reduction has in fact occurred.

Introducing a third readout $M_3$, with its measurement basis orthogonal to that of $M_2$ ($\theta_2=\pi/2$), produces a joint density matrix for the three measurements that deviates from the prediction of standard wavefunction collapse~\cite{GlickAdami2020}.

The unitary formalism predicts:
\begin{widetext} \be \rho_{123}\label{joint} = \frac{1}{4} \begin{pmatrix} \cos^2(\phi)\mathbbm{1} & -\cos^2(\phi)\sigma_z & 0 & 0 \\ -\cos^2(\phi)\sigma_z & \cos^2(\phi)\mathbbm{1} & 0 & 0 \\ 0 & 0 & \sin^2(\phi)\mathbbm{1} & \sin^2(\phi)\sigma_z \\ 0 & 0 & \sin^2(\phi)\sigma_z &\sin^2(\phi) \mathbbm{1} \end{pmatrix}\;,\ \ \ \ \ \ee \end{widetext}
where $\mathbbm{1}$ denotes the identity matrix and $\sigma_z$ the third Pauli matrix.
This matrix is incompatible with a collapse picture because it contains off-diagonal elements. These terms imply coherence among the measurements, which can, in principle, be revealed via interference. The corresponding purity,
\begin{equation}
C(\rho_{123}) = \mathrm{Tr}(\rho_{123}^2) = \tfrac{1}{4}\bigl(1 + \cos^2(2\phi)\bigr),
\label{purity}
\end{equation}
quantifies this residual coherence.

In contrast, the standard collapse theory predicts a fully diagonal joint state,
\begin{widetext} \be \rho_{\rm coll_{123}}\label{rho_coll} = \frac{1}{4} \begin{pmatrix} \cos^2(\phi)\mathbbm{1} & 0 & 0 & 0 \\ 0 & \cos^2(\phi)\mathbbm{1} & 0 & 0 \\ 0 & 0 & \sin^2(\phi)\mathbbm{1} & 0 \\ 0 & 0 & 0 &\sin^2(\phi) \mathbbm{1} \end{pmatrix}\;,\ \ \ \ \ \ee \end{widetext}
with purity
\begin{equation}
C(\rho_{\mathrm{coll_{123}}}) = \mathrm{Tr}(\rho_{\mathrm{coll}_{123}}^2)
= \tfrac{1}{8}\bigl(1 + \cos^2(2\phi)\bigr).
\label{C_rho_coll}
\end{equation}
The purity predicted by the unitary theory is therefore twice that of the collapse theory, an experimentally testable distinction explored later. 

While decoherence, i.e., the loss of quantum coherence through interactions with uncontrolled environmental degrees of freedom~\cite{zeh1970_decoherence,joos2013decoherence-book,RMP_Zurek_Decoherence2003,schlosshauer2007decoherence,SCHLOSSHAUER2019_decoherence,deleglise2008decoherence,brune1996decoherence,kiefer2022quantum_decoherence,broome2010quantumwalksdecoherence,kwiat2000decoherence-freesubspaces,RMP2005_Decoherence_measurementproblem_Schlosshauer,PRL_Decoherence-Free_HW,Joos_arxiv_decoherence,stanford_RoleofDecoherence}, will ultimately suppress the off-diagonal elements of $\rho_{123}$, thereby transforming it into $\rho_{\mathrm{coll}_{123}}$, the coherence present in Eq.~(\ref{joint}) can be revealed through interference if decoherence is sufficiently prevented. Moreover, the decoherence of specific measurements within the measurement chain can be deliberately controlled. For example, decohering only the second readout $M_2$ allows one to verify that the resulting state $\rho_{\mathrm{coll}_{123}}$ and its purity $C(\rho_{\mathrm{coll}_{123}})$ agree with the predictions of the standard collapse picture.

Although decoherence can affect any quantum system, the unitary model described here is not an environment-induced decoherence model of quantum measurement. Unlike those approaches, it does not rely on coupling to an external macroscopic bath to absorb quantum phases; rather, the coherence and its possible loss arise entirely within the closed system of sequential measurements.

\begin{figure}[]  
  \centering
  \includegraphics[width=1.0\columnwidth]{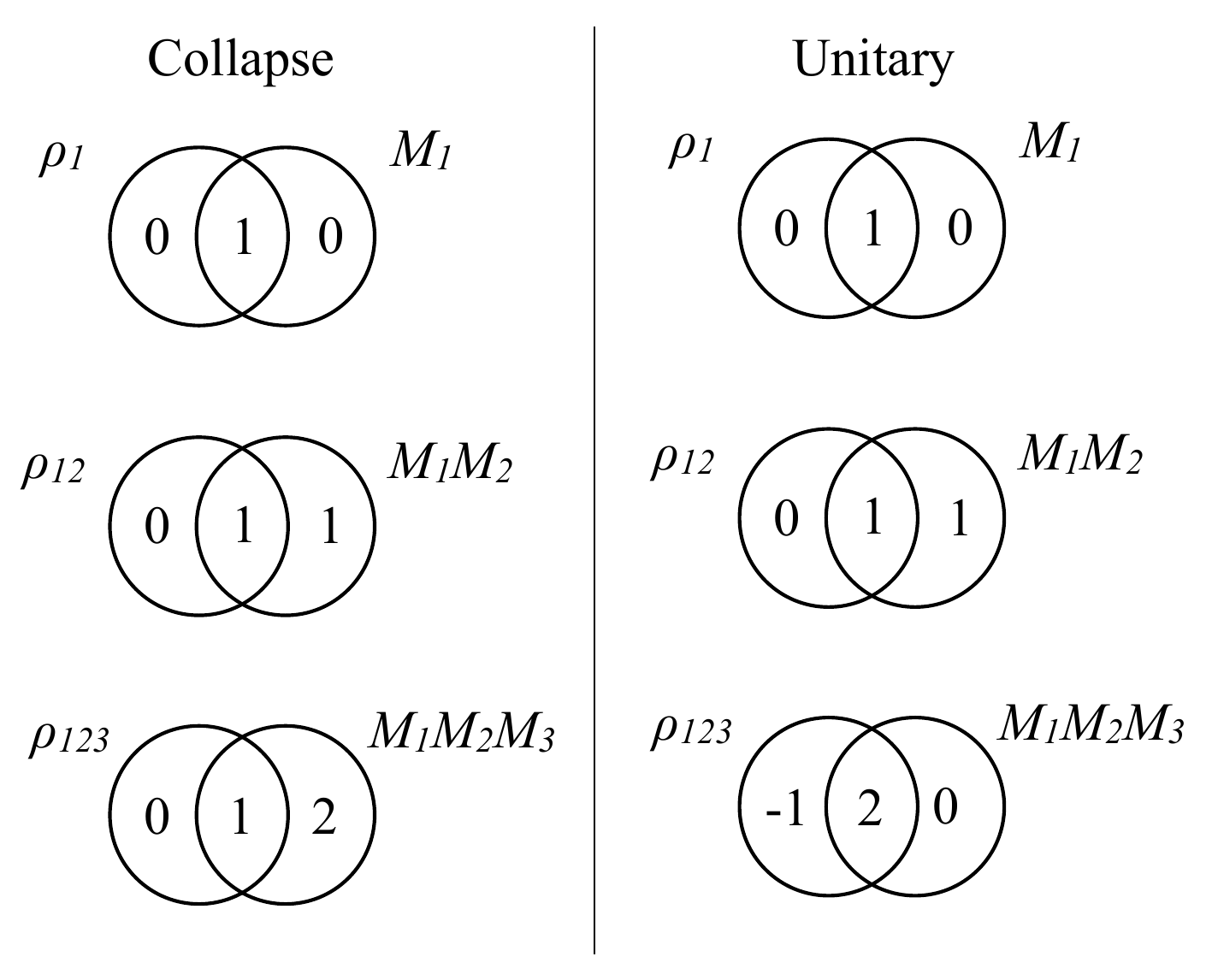} 
  \caption{Information-theoretic Venn diagrams showing conditional and shared entropies between the quantum system $Q$ and the measurement systems after one, two, and three consecutive measurements. The left column corresponds to the standard collapse theory and the right column to the unitary theory. Three systems $M_1$, $M_2$, and $M_3$ measure a fully mixed quantum system ($\phi=\pi/4$) with relative angles $\theta_1 = \theta_2 = \pi/4$. In the unitary case, the third measurement creates entanglement between $Q$ and the collective measurement system, signaled by negative conditional quantum entropies.}
  \label{fig:QVD}
\end{figure}

Even though coherence between measurements does not reveal the quantum state directly, it provides indirect evidence about the quantum state's post-measurement character. Figure~\ref{fig:QVD} illustrates this connection using information-theoretic Venn diagrams that depict the conditional and shared entropy between the quantum system $Q$ and the measurement system after one, two, and three consecutive measurements of a fully mixed state ($\phi = \pi/4$) with relative measurement angles $\theta_1 = \theta_2 = \pi/4$. For one and two consecutive measurements, the diagrams are identical for both theories: in each case, the measurement systems collectively acquire one bit of information from the initial fully mixed state. In the collapse picture, the third measurement yields no additional information beyond the second. In contrast, the unitary theory predicts that after the third measurement, the three systems together become entangled, and therefore in superposition, with the quantum state. The presence of this entanglement implies that unrealized amplitudes must have persisted after the first and second measurements as well. Negative conditional quantum entropies serve as indicators of entanglement in such composite systems~\cite{Cerf&Adami1997,Cerf&Adami1999,wilde2013quantum,horodecki2005partial}.

Furthermore, for the sake of clarity, it is important to distinguish between two terms: measurements and detectors. A measurement is the process of ascribing a defined quantum state to a system (or particle) of interest. In this sense, one can discuss a series of ``sequential Stern-Gerlach experiments'' as presented by Sakurai and Napolitano~\cite{sakuraimodernQM}, and in our case, a series of measurements denoted by $M_1$, $M_2$, and  $M_3$ in Fig.~\ref{fig:setup}. As for detectors, they are the devices that provide the final readout.

In order to test the description of three consecutive measurements in the unitary theory, it is imperative that the quantum system is protected from interacting with any other system besides the selected measurement systems. The reason for this is clear: any interaction not registered within measurement devices is akin to an unobserved measurement that we would need to average over, and will destroy the coherence predicted by the theory. This is why photonics is an ideal platform for such experiments, since photons hardly interact with the environment, leading to negligible environmental decoherence. In this way, the quantum state of a photon is protected from environmental decoherence and can easily be measured using stable decoherence-free pointer states in the form of which-path information. If an experiment would be carried out on other platforms, for example superconducting qubits~\cite{superconductingqubits2019_Krantz,devoret2013superconducting,minev2019catch}, ion traps~\cite{RMP_wineland2003,postler2022blatt}, or neutral atoms~\cite{briegel2000neutralatoms,Henriet2020neutralatoms,briegel2000neutralatoms}, it would be crucial that consecutive measurements are carried out fast enough to prevent environmental decoherence to occur, i.e., it would be necessary to prevent the occurrence of intermediate unobserved measurements.

If the pointer states were to be amplified, for example via a CCD readout, the coherence between pointer states predicted by our theory would be lost. It is not the amplification of the first or third pointer state that would destroy coherence, since these are already incoherent. It is the amplification of the pointer state of the middle measurement that would destroy coherence (Eq.~(4.8))~\cite{GlickAdami2020}. For this reason, we carry out all three measurements in succession, store the information in three pointer which-path variables, and then, rather than amplify those, we perform a direct measurement of the joint density matrix of the three measurement systems via classical readouts (see note~\footnote{In Ref.~\cite{Dicke1989}, Dicke has argued that before the readout, the measurement has already occurred.}). In this manner, we can test whether it is the readout itself that destroys coherence (via collapsing the quantum wavefunction, i.e., a detector-induced decoherence), or whether environment-induced decoherence is the culprit~\cite{RMP_Zurek_Decoherence2003}. If we were to simply amplify the which-path pointers, this distinction could not be made.

\section{Experiment}

\begin{figure*}[]   
  \centering
   \includegraphics[width=1.0\textwidth]{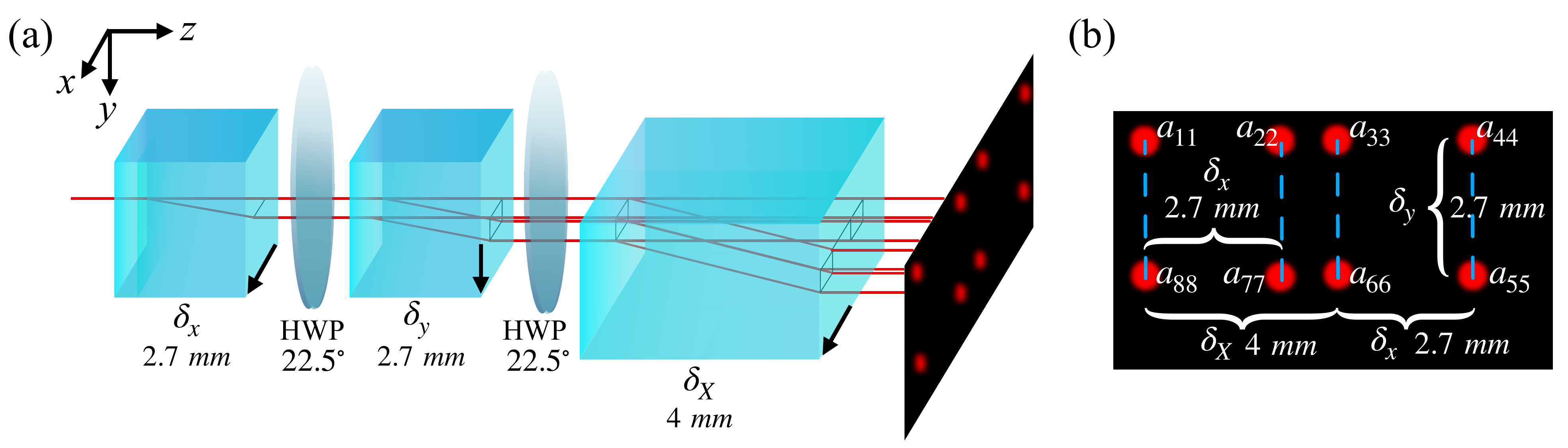}
   \caption{(a) Conceptual view of three consecutive measurements implementing a sequence of non-commuting measurements ($\pmb{\pi}_H\pmb{\pi}_D\pmb{\pi}_H$) using calcites as beam displacer and half-waveplates. (b) The final geometry for the eight beam paths. The dashed blue lines indicate coherence according to the unitary theory.}
   \label{fig:conceptual}
\end{figure*}

\begin{figure*}[] %
   \centering
   \includegraphics[width=1.0\textwidth]{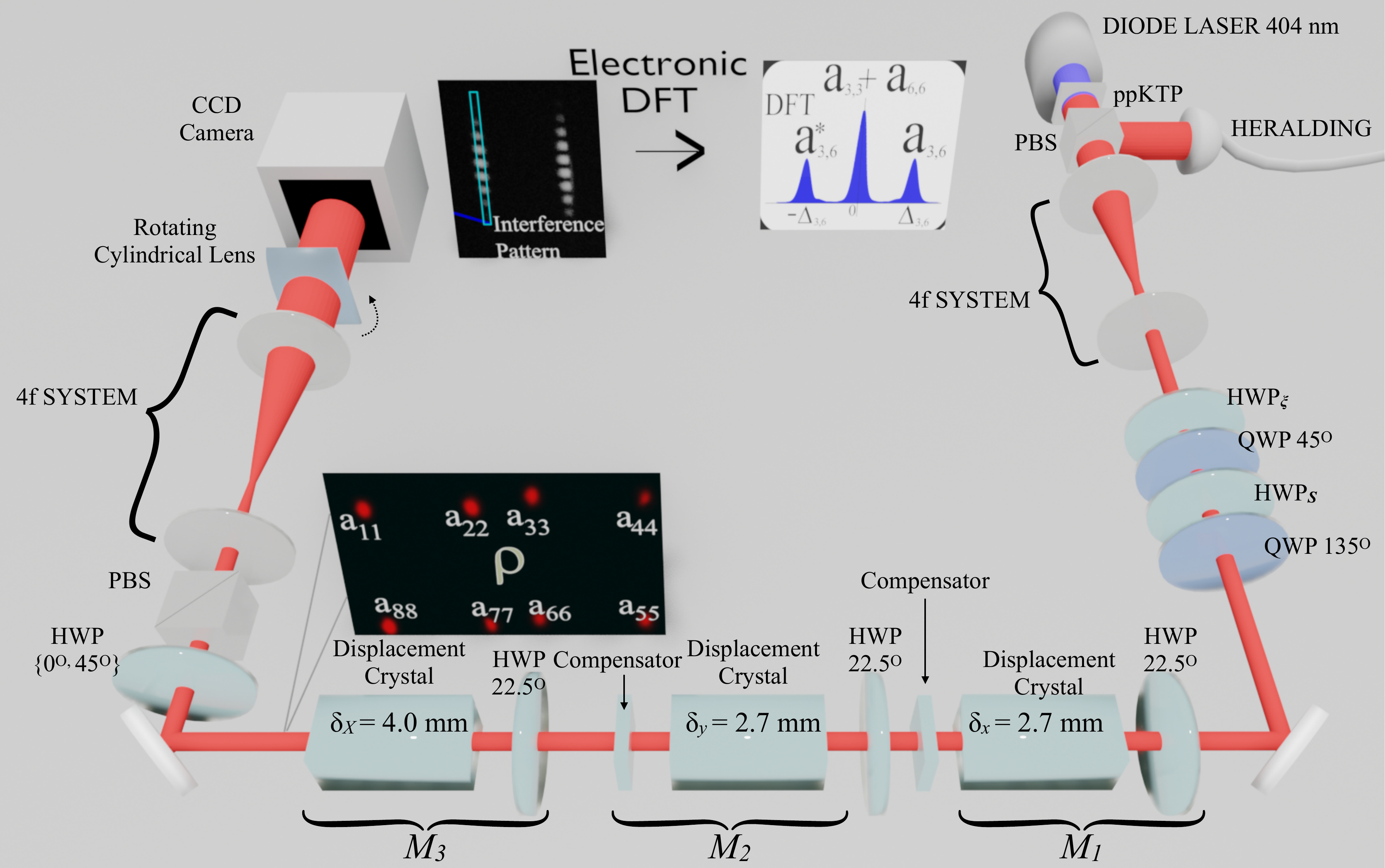} 
   \caption{State preparation, sequential measurements, and quantum state reconstruction. Photons are generated by the SPDC process. A 4$f$-system is built to ensure all the optics are within the Rayleigh length. A combination of half-wave plates (HWP) and quarter-wave plates (QWP) prepares a mixed state, while the HWP$_S$ spins faster than the collection time of the camera. Next, a series of measurements is implemented using calcite displacement crystals (named $\delta_x$, $\delta_y$, and $\delta_X$). Lastly, another 4$f$-system ($f_1 = 1000$ mm and $f_2 = 400$ mm) is used to image the eight path modes in an electron-multiplying CCD (EMCCD) camera. Optical Fourier transform (OFT) along the OFT-axis is performed by rotating a 250~mm cylindrical lens. The final coherence readout is obtained by Discrete Fourier Transform (DFT).}
   \label{fig:setup}
\end{figure*}

Figure 3(a) illustrates the sequence of three measurements implemented at relative angles $\theta_1=\theta_2=\pi/4$ in polarization space, sketched in Fig.~1. Because the polarization rotation matrix involves $\cos(\theta/2)$, these settings correspond to orthogonal measurement bases. The diagram also shows the eight possible outcomes corresponding to the three binary measurements.

In this quantum-optical setup, beam-displacer crystals shift a polarized beam transversely according to its polarization. This approach is experimentally convenient for two reasons. First, it enables non-destructive measurements, in which the photon continues to propagate after each read-out, allowing multiple sequential measurements on the same quantum state. Second, the parallel propagation of the beams ensures that environmental fluctuations are largely common-mode, thereby preserving any residual coherence among the paths.

Although non-destructive polarization measurements are often implemented in the weak-measurement regime using small beam displacements~\cite{Lundeenetal2011,LundeenBamber2012,martinez2021JWM}, in our experiment we use sufficiently large displacements so that the measurements are fully projective, i.e., strong measurements~\cite{ValloneDequal2016,jordan&siddiqiBook}.

The first beam displacer crystal shifts horizontally polarized light by $\delta_x=2.7$ mm in the $x$-direction, while vertically polarized light remains unshifted. This conditional translation implements the measurement operator $M=e^{i\delta_x\pmb{\pi}_H\otimes \pmb{p_{k}}}$ where $\pmb{p}_k$ is the transverse-momentum operator along direction $k\in{x,y}$ (up to normalization), and $\pmb{\pi}_H=\ket{H}\!\bra{H}$ projects onto horizontally polarized states.

The second beam displacer shifts vertically polarized light by $\delta_y=2.7$ mm along $y$. Half-wave plates (HWP) set at $22.5^{\circ}$ are placed before and after the beam displacer crystal, so that the combined action realizes a measurement in the diagonal basis ($\pmb{\pi}_D$), corresponding to an equal superposition of horizontal and vertical polarizations.

Finally, the third beam displacer crystal shifts horizontally polarized light by $\delta_X=4$ mm in the $x$-direction, implementing the final $\pmb{\pi}_H$ measurement. A summary of all displacements is given in Table \ref{tab:shifts}. The output geometry of the eight resulting beams is shown in Fig.~\ref{fig:conceptual}(b); dashed blue lines indicate the coherent paths predicted by the unitary theory.

\begin{table}[htbp]
    \centering
     \topcaption{Summary of physical shifts for each of the eight paths shown in Fig.~\ref{fig:conceptual} (a).}
\begin{tabular}{ c | c } 
        $s_x,s_y$ & $s_x,s_y$  \\
       \hline
          $0,0$      &   $\delta_x,\delta_y$\\
        $ \delta_x,0$  &   $\delta_X,\delta_y$ \\
          $\delta_X,0$  & $\delta_X+\delta_x,0$  \\
          $0,\delta_y$   &  $\delta_X+\delta_x,\delta_y$ \\
       \hline
    \end{tabular}
    \label{tab:shifts}
 \end{table}
 
Figure 4 shows the experimental setup used to implement the three sequential polarization measurements. We used a periodically poled KTP (ppKTP) crystal in a Spontaneous Parametric Down-Conversion (SPDC) source~\cite{Christ2013SPDC}, pumped by a diode laser at $\lambda= 404$ nm and producing photon pairs at $\lambda= 808$ nm (details in Appendix B). One photon of the pair heralds its twin, a single polarized photon that we use as our quantum system. The heralded photon first undergoes a state-preparation stage consisting of a beam expander and a set of half-wave plates (HWP) and quarter-wave plates (QWP). A mixed input state is produced by rapidly rotating HWP$_S$ at a rate faster than the camera integration time. A sequence of beam displacer crystals (labeled $\delta_x$, $\delta_y$, and $\delta_X$) together with HWP set at 22.5° then implements the non-commuting measurements $\pmb{\pi}_H\pmb{\pi}_D\pmb{\pi}_H$. In certain configurations, walk-off compensation crystals are inserted to counteract birefringence \cite{Kwiat2009,kwiat2000decoherence-freesubspaces}, providing a controllable way to introduce decoherence.

This arrangement produces co-propagating beams that enable the double-slit–type interference required for the state reconstruction. Because the path modes share all optical elements and remain spatially close, the setup maintains excellent phase stability and minimizes vibration noise.

To measure the joint density matrix after three consecutive measurements on a prepared quantum state $\Tilde{\rho}_1$ (Eq.~\ref{rho_1_tilde}), we vary the first HWP in Fig.~\ref{fig:setup} from $\xi=0^{\circ}$ to $\xi=22.5^{\circ}$ in 2.25° increments. This adjustment varies the mixing angle from $\phi=0$ (pure state) to $\phi=\pi/4$ (fully mixed), with $\phi=2\xi$. The first measurement defines the reference basis, while the second and third are oriented at $\theta_1=\theta_2=\pi/4$ relative to the preceding bases, by setting the HWP of $M_2$ and $M_3$ to 22.5°. This configuration therefore implements the sequence $\pmb{\pi}_H\pmb{\pi}_D\pmb{\pi}_H$\footnote{An equivalent configuration can be realized by keeping the initial state fixed and decreasing the mixing angle from $\phi=45^{\circ}$ to $0^{\circ}$ rather than rotating the initial state.}.

We also perform a second experiment in which birefringence is not compensated after the second beam displacer crystal, thereby introducing decoherence in the second measurement stage. According to our theory, the reconstructed density matrix in this configuration should correspond to that predicted by the standard collapse theory.

We reconstruct the quantum state using the interference-based tomography techniques described in Appendix C. This stage recovers the matrix elements of Eq.~(\ref{joint}) through interference among the output paths. 

Direct observation of destructive interference can be ambiguous for small angles, where the difference between the absence of interference and near-perfect destructive interference becomes hard to discern. To avoid this ambiguity, we perform all measurements using an initial quantum state that is non-diagonal, prepared by measuring in the diagonal basis ${D,\bar D}$ rather than the horizontal–vertical basis ${H,V}$ (see Appendix E). In this configuration, coherence manifests as constructive interference, which is easier to distinguish experimentally from the incoherent case predicted by the collapse theory. For such a 
 {\em non-diagonal} initial state
\begin{equation}
{{\Tilde{\rho}_{1}}} = \frac{1}{2}
\begin{pmatrix}
1 && \cos{2 \phi} \\
\cos{2 \phi}  && 1
\end{pmatrix}\;,\label{rho_1_tilde}
\end{equation}
the density matrix after the second measurement becomes (see Appendix~\ref{Approt}):
\begin{widetext}
\begin{equation}
\tilde\rho_{12}=\frac14
   \begin{pmatrix}
    1 &  \cos(2\phi) & 0 & 0 \\
      \cos(2\phi)& 1& 0 & 0 \\
      0 & 0 & 1 & -\cos(2\phi)\\
      0 & 0 & -\cos(2\phi) & 1\\
    \end{pmatrix}, \label{tilderho_12}
\end{equation}
\end{widetext}
with purity $C(\tilde\rho_{12})=\frac14(1+\cos^2(2\phi))$, consistent with the prediction of the collapse model. After the third measurement stage, the density matrix becomes (with $c=\cos(2\phi)$; see Appendix~\ref{Approt} for the derivation)
\begin{equation}
\Tilde{\rho}_{123}=\frac18
\begin{pmatrix} 
1 & 1 & c & -c & 0 & 0 & 0 & 0 \\
1 & 1 & c & -c & 0 & 0 & 0 & 0 \\
c & c & 1 & -1 & 0 & 0 & 0 & 0 \\
-c & -c & -1 & 1 & 0 & 0 & 0 & 0 \\
0 & 0 & 0 & 0 & 1 & -1 & c & c \\
0 & 0 & 0 & 0 & -1 & 1 & -c & -c \\
0 & 0 & 0 & 0 & c & -c & 1 & 1 \\
0 & 0 & 0 & 0 & c & -c & 1 & 1 \\
\end{pmatrix},\label{rho_123_tilde}
\end{equation}
with purity
\begin{equation}   C(\Tilde{\rho})_{123}=\tr(\Tilde{\rho}_{123}^2)=\frac14(1+\cos^2(2\phi))\;,\label{purity_rho_123_tilde}
\end{equation}
exactly as Eq.~(\ref{purity}). Thus, although using the non-diagonal initial state alters the detailed form of the post-measurement density matrices, the overall purity, and therefore the measurable coherence, remains unchanged. Furthermore, while measurements on the rotated initial state can lead to joint density matrices that are non-diagonal in both collapse-based and unitary theories, their elements are significantly different at every angle.

Another way to understand our experimental procedure is as follows. A quantum measurement is a two-step process: a latent measurement puts the measurement results into qubit pointers, followed by amplification of these pointers, and the amplification destroys the coherence between these pointers. In fact, the amplification of qubits with macroscopic devices causes any coherence between these qubits to be lost. As such, a reader could argue that one should not observe any coherence after the readout of our measurement, since the readout devices are classical objects. However, rather than amplifying each of the three pointer states, in our experiment we measure the density matrix of the pointer states via interference, not an amplification of the pointer states. In other words, our experiment is designed to measure the relative phases, not the which-path information. One could never perform this experiment via standard amplification of counts in CCD detectors, and thus, one solution is to measure the density matrix before the amplification process.

\section{Results}

\begin{figure*}[]   
   \centering
   \includegraphics[width=1.0 \textwidth]{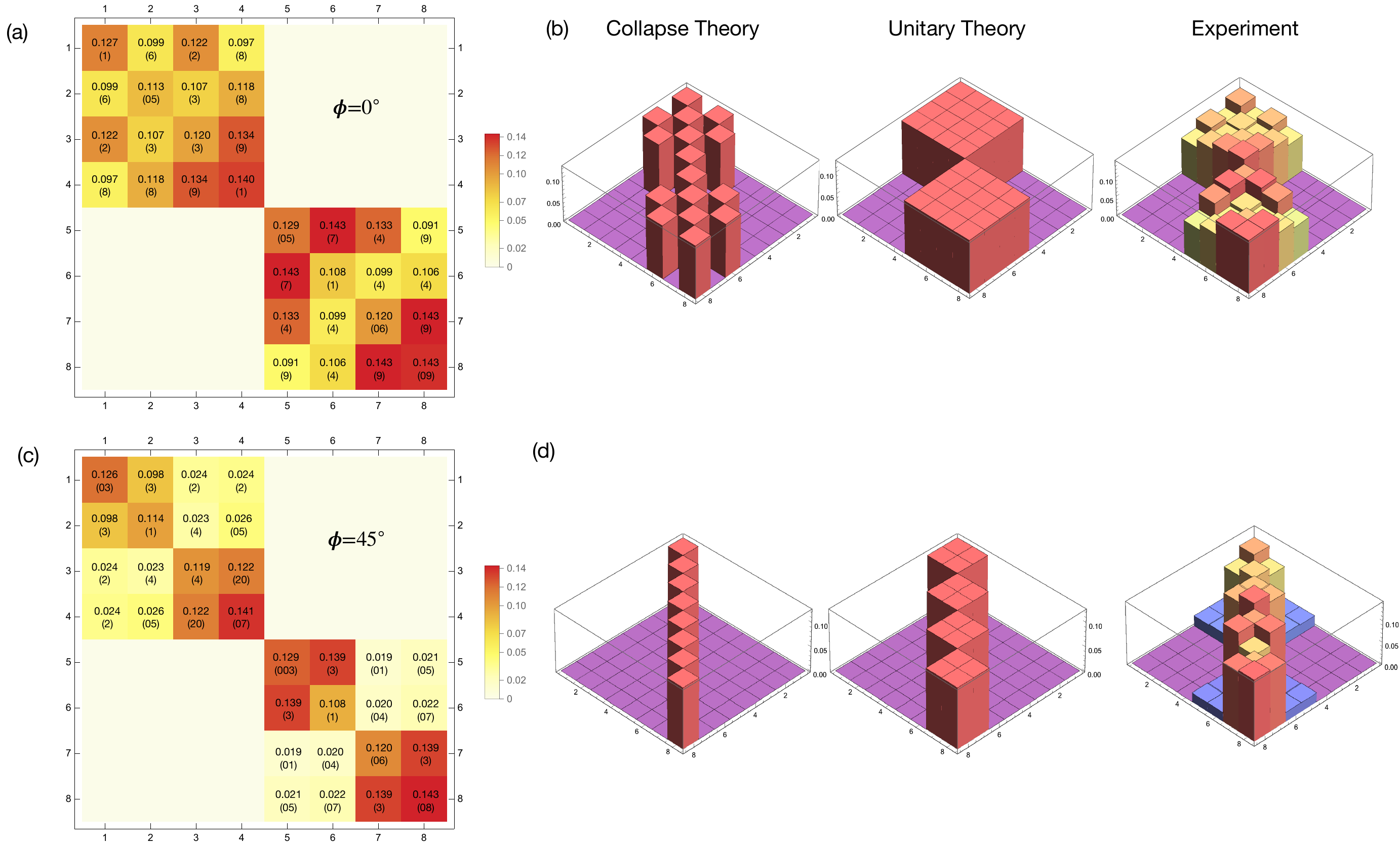} 
   \caption{ (a) Reconstructed joint density matrix [Eq.~(\ref{rho_123_tilde})] for a completely mixed input state ($\phi=0$). Each square represents the absolute value of a matrix element after the third measurement, averaged over multiple experimental runs; one standard deviation is indicated numerically within each square. The color scale (right) encodes the mean amplitude. The upper-left and lower-right $4\times4$ blocks correspond to the populated subspaces, while the off-diagonal $4\times4$ blocks are zero in both theories.
(b) Predicted absolute values of the joint density matrix for $\phi=0$ under the collapse model, the unitary model, and the experiment (right), shown side-by-side as Manhattan plots. Experimental amplitudes correspond to those reported numerically in panel (a).
(c) Reconstructed joint density matrix for a diagonal (pure) input state, $\phi=45^{\circ}$, displayed as in (a).
(d) Predicted absolute values of the joint density matrix for $\phi=45^{\circ}$ under the collapse, unitary, and experimental results, shown as in (b). Additional data for intermediate values of $\phi$ are presented in Appendix \ref{extramat}. In all cases, including those shown here, the experimental results agree with the unitary theory and disagree with the standard collapse theory.
   \label{fig:density8x8}
   }
\end{figure*}

\begin{figure}[]  
   \centering
   \includegraphics[width=1.0\columnwidth]{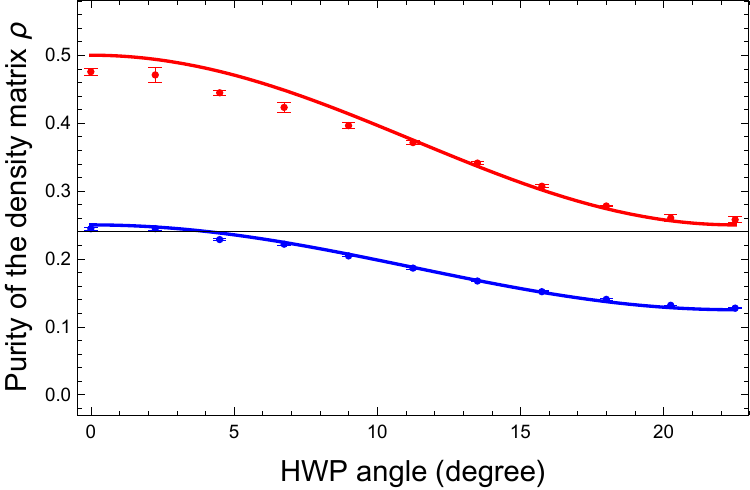} 
   \caption{Purity of the reconstructed joint density matrix after three measurements as a function of the half-wave-plate (HWP) angle $\xi$, with $\phi=2\xi$; thus, $\xi=0^{\circ}$ corresponds to a pure initial state and $\xi=22.5^{\circ}$ to a fully mixed state. Points represent averages over independent experimental runs (see Table II); error bars denote standard errors. Solid lines show the corresponding theoretical predictions. Red circles correspond to experiments in which walk-off compensation was applied after the first and second crystals, preserving coherence. Their measured purities agree with the unitary-theory prediction [Eq.~(\ref{purity_rho_123_tilde})] and are incompatible with the standard collapse model. Blue circles correspond to experiments performed without walk-off compensation after the second crystal, introducing controlled decoherence. Their measured purities follow the prediction of Eq.~(\ref{C_rho_coll}), consistent with the collapse model. This behavior is, however, also predicted by the unitary theory when decoherence is explicitly included. 
   }
   \label{fig:Purity-density}
\end{figure}   

Figure \ref{fig:density8x8} shows the reconstructed joint density matrices after three consecutive measurements for initial states with $\phi=0$ (Fig.~\ref{fig:density8x8}(a) and Fig.~\ref{fig:density8x8}(b)) and $\phi=\pi/4$ (Fig.~\ref{fig:density8x8}(c) and Fig.~\ref{fig:density8x8}(d)), averaged over independent experimental runs with walk-off compensation applied after crystals 1 and 2. Figures ~\ref{fig:density8x8} (a) and (c) display the experimentally reconstructed absolute values of the matrix elements of Eq.~(\ref{rho_123_tilde}), arranged in $4\times4$ submatrices; elements not shown are theoretically zero in both the collapse and unitary descriptions. Figures~\ref{fig:density8x8} (b) and (d) compare the corresponding theoretical predictions for the collapse and unitary theories with the experimental results. While some matrix elements expected to be zero show a nonzero value due to intrinsic background and detection noise, these elements are at least one order of magnitude smaller than the nonzero elements predicted by theory, which makes them clearly distinguishable. Additional reconstructed matrices for intermediate values of $\phi$ are presented in Appendix \ref{extramat}.

In all cases, the reconstructed density matrices after three consecutive measurements agree quantitatively with the predictions of the unitary theory and are incompatible with the collapse theory. This agreement implies that the unrealized amplitudes of the initial quantum state persist after the second measurement. Any decoherence of the middle measurement merely renders this existing coherence temporarily unobservable, without eliminating it.

To further test whether the density matrix conforms to the unitary theory, we calculate the purity of the state shown in Eq.~(\ref{rho_123_tilde}) (the purity is the same as Eq.~(\ref{purity}), as it does not depend on the initial state) from the reconstructed matrix at all angles, and compare to the predicted purity in the standard collapse theory as Eq.~(\ref{C_rho_coll}). The results are presented in Fig.~\ref{fig:Purity-density}, where data are shown as points with standard error, and the solid lines are the theoretical predictions. The red curve corresponds to three consecutive measurements where crystals 1 and 2 are compensated for walk-off, i.e., no decoherence. In this case, the data are compatible with the unitary theory and support the no-collapse view. The blue curve corresponds to three consecutive measurements with walk-off of the 1$^{\text{st}}$ crystal compensated, but walk-off of the 2$^{\text{nd}}$ crystal not compensated, thus allowing for decoherence. In this case, the data are compatible with the collapse picture, and show that the standard model of quantum measurement is only recovered when decoherence is explicitly allowed or introduced. 

\section{Discussion}

The interpretation of quantum measurement has been debated for nearly a century, and numerous frameworks have been proposed to make sense of quantum theory~\cite{Lombardietal2017,laloe2019we,auletta2001foundations,cabello2017interpretations}. Broadly, these approaches fall into two categories: those that assume a physical reduction (collapse) of the wavefunction upon measurement, and those that maintain strictly unitary evolution throughout. 

Here, we have tested a specific prediction of a unitary theory of measurement~\cite{GlickAdami2020}: that three or more consecutive measurements on the same quantum system can reveal amplitudes that would be regarded as “collapsed” under the standard picture. This theory predicts that the joint density matrix of the measurement system contains off-diagonal elements arising from coherence in the intermediate measurement, even though such coherence is absent in the density matrix after the second measurement alone. Our reconstruction of the joint density matrix from repeated single-photon measurements using direct quantum state tomography confirms this prediction. The experimentally observed nonzero off-diagonal terms are consistent with the unitary description and incompatible with the collapse assumption, central to the Copenhagen interpretation of quantum mechanics.

The unitary theory resolves several long-standing conceptual difficulties in the interpretation of quantum mechanics. A central question, dating back to the Einstein–Bohr debate~\cite{Einsteinetal1935,Bohr1935}, concerns the question: Do physical systems possess definite properties prior to, and independent of, measurement? As commonly posed, this question implicitly assumes that the post-measurement state of the apparatus accurately reflects the pre-measurement state of the system. The assumption of such perfect correlation between system and device is pervasive in discussions of measurement~\cite{Wigner1963}, yet it conflicts with a fundamental quantum information constraint: the no-cloning theorem~\cite{Dieks1982,WoottersZurek1982}.

In the unitary framework, the measurement device cannot, in general, perfectly represent the properties of the system it probes, except in the special case where the system was prepared in the same basis as the measurement device. In the opposite, orthogonal case, the measurement outcome carries no information at all about the system’s prior state. Consequently, after measurement, the quantum system {\em itself} retains information about the “unrealized” quantum amplitudes of the wavefunction not captured by the detector, which act as a kind of memory of prior possibilities.

This perspective requires us to abandon the notion that detector readouts directly reveal the underlying physical state of a quantum object. Instead, measurement outcomes must be understood as statistical data from which we can infer how identically prepared systems will interact with the same class of measurement devices. In practice, this inference emerges only after repeated trials in accordance with Born’s rule.

The Copenhagen interpretation remains the most widely held view of quantum measurement among physicists~\cite{schlosshauer2013snapshot,Zinkernagel2016,Henderson2010,Nature2025survey}, and the question of which interpretation best reflects physical reality is often regarded as a matter of personal preference rather than empirical testability. Within this framework, the collapse of the wavefunction is commonly treated as one of its defining principles~\cite{auletta2001foundations,Omnes-InterpretationQM94,Faye-copenhagen-stanford2024} . One aim of the present work is to shift this long-standing discussion from interpretational preference to experimental verification.

The conclusions reported here apply broadly to any interpretation that, like the Copenhagen view, treats wavefunction collapse as a physical erasure of unrealized amplitudes. Our results show that when decoherence is prevented, the system evolves coherently in accordance with the unitary theory, whereas collapse-like behavior emerges only when decoherence is allowed. In this sense, what is conventionally attributed to “collapse” can be understood as the empirical signature of decoherence within a fully unitary framework. Moreover, the conclusions presented here also hold for any interpretation of quantum mechanics that shares the same view of the collapse of the wavefunction as the Copenhagen interpretation, or at least that views the collapse of the wavefunction as an erasing mechanism.

We note that the term ``collapse models'' is sometimes used more narrowly to denote a subset of interpretations of quantum mechanics, namely, objective-collapse theories, such as the Ghirardi–Rimini–Weber model and Penrose’s gravitational collapse proposal~\cite{Stanford-collapse,WayneQuantumMeas,RMPcollapse2013}. The present results do not directly test these specific mechanisms, but rather the general assumption shared by many interpretations, i.e., that measurement involves a fundamental non-unitary reduction of the quantum state. Moreover, the experimental distinction between unitary evolution and decoherence contributes to a deeper understanding of decoherence itself, a central issue in quantum information science that underlies numerous applications, including quantum computing and quantum optomechanics~\cite{RMP_wineland2003,haroche2006book,ladd2010review,superconductingqubits2019_Krantz,devoret2013superconducting,RMPoptomech,meystre_qoptomech,schumacher2010book,jordan&siddiqiBook}.

The results presented here also bear on the issue of the time symmetry of quantum measurement. In 1964, Aharonov, Bergmann, and Lebowitz (ABL) demonstrated that, for appropriately constructed measurement sequences, knowledge of both past and future outcomes allows the outcome of an intermediate measurement to be inferred with certainty~\cite{Aharonovetal1964}. Their analysis highlighted an apparent asymmetry in the measurement process: while past measurements constrain future outcomes, the converse seemed less evident. Our findings suggest that this asymmetry is only superficial. The coherence among consecutive measurements observed here indicates that, within a fully unitary evolution, information about unrealized amplitudes connects past and future measurement events in a single time-symmetric framework. In this view, the correlations encoded in the joint density matrix embody the same principle envisioned by ABL, not as a special case requiring contrived post-selection but as a general feature of all quantum measurement chains.

As we commented in section~\ref{sec:theory}, to observe any manifestation of unrealized quantum amplitudes, i.e., the quantum amplitudes of a superposition that are not reflected in the readout of a measurement device, it is imperative to search for their effects before decoherence prevails. Decoherence can occur either when the quantum system of interest itself undergoes decoherence (i.e., it interacts with an uncontrolled degree of freedom, which we call environment-induced decoherence), or when the joint entangled system composed of the quantum system of interest and the measuring apparatus decohere, a process we call detector-induced decoherence.

Some additional distinctions between unitary and non-unitary theories of quantum measurement need to be addressed. In theories in which a measurement involves a non-unitary process that extinguishes the off-diagonal elements of the density matrix of the quantum system (or, as in Quantum Darwinism, which relegates them to an inaccessible abstract space~\cite{Zurek2007,Zurek2009}), the off-diagonal elements are lost forever (see note~\footnote{In Ref.~\cite{dirac1981principles-book}, Dirac states: ``...a measurement always causes the system to jump into an eigenstate of the dynamical variable that is being measured...''}). This is not the case in the unitary theory, provided one observes the caveats imposed by decoherence (since in the unitary theory environment-induced decoherence can occur just as readily, as discussed earlier). Moreover, it is unnecessary to assume that a quantum system has jumped into an eigenstate (erasing all other amplitudes) just because the detector takes on state $i$, because while Born's rule correctly quantifies this probability as $|\psi_i|^2$, it is erroneous to deduce that the quantum state itself collapsed into state $\psi_i$ (as Born concluded~\cite{Born1926}). The unitary theory implies that in the worst case the quantum state of interest and the detector state can be completely uncorrelated, as is shown in the worst case for example angle $\phi=45^\circ$ in our experiment.

As the quantum system of interest interacts with detectors, many decoherence channels may become available when a detector's marginal density matrix contains off-diagonal elements.  Thus, decoherence, through either way (detector-induced or environmental-induced), is the mechanism responsible for the quantum-to-classical transition in the sense that decoherence correctly accounts for the suppression of macroscopic superpositions. Any classical macroscopic object has already undergone decoherence, and thus, no classical macroscopic object can be a true reflection of a quantum system. Therefore, no classical object used as a detector can perfectly reflect the quantum state that is supposed to measure.

While in our experiment we have explicitly prevented the decoherence of the middle measurement (by encoding its state in the which-path information), the unitary theory implies that even a classical detector would be found in a superposition after the third measurement (see Eq.~(3.11) in Ref.~\cite{GlickAdami2020}, where each detector is written in terms of $n$ correlated bits, where $n$ is arbitrarily large). Such a detector would, of course, decohere immediately after the third measurement.  Whether such a hypothetical ``recoherence-decoherence'' sequence for a classical ``middle measurement'' can be experimentally observed is an open question. 

\section{Conclusion}

We have experimentally tested two fundamentally different descriptions of quantum measurement: the collapse theory, commonly associated with the Copenhagen interpretation, and the unitary theory, in which the measurement process is entirely coherent. Using single photons and encoding information in polarization as well as path degrees of freedom, we performed three consecutive measurements on the same quantum system. Two key pieces of evidence: the presence of nonzero off-diagonal elements in the reconstructed joint density matrix and the measured purity of that matrix, disagree with the predictions of the collapse picture and are consistent with the unitary theory. Collapse-like behavior was recovered only when decoherence was deliberately introduced, underscoring the central role of decoherence in producing the appearance of wavefunction collapse.

These results provide a direct, experimentally testable distinction between collapse-based and fully unitary formulations of quantum measurement. They extend to any interpretation of quantum mechanics that invokes collapse as a physical erasure mechanism, and they demonstrate that the apparent reduction of the wavefunction can be understood as an emergent consequence of decoherence within an otherwise unitary framework. These findings also suggest that quantum measurement, viewed as a fully unitary process, possesses an inherent time symmetry linking past and future measurement events through the persistent coherence of unrealized amplitudes. 

This work revisits one of the foundational pillars of quantum mechanics, the wavefunction collapse, to address this controversy in a testable way and clarifies important features of quantum measurements.

\section{Acknowledgment}
We acknowledge D. Curic for experimental support and help with the manuscript, and J. R. Glick for theoretical support. We also thank A. C. Martinez-Becerril and F. C. Cruz for help with figures. This work was supported by the Canada Research Chairs (CRC) Program, the Natural Sciences and Engineering Research Council (NSERC), the Canada Excellence Research Chairs (CERC) Program, the Canada First Research Excellence Fund award on Transformative Quantum Technologies.

\clearpage
\appendix
\begin{widetext}
\section*{Appendices}
\label{Supp}

\section{Creation of mixed states}
To create mixed states from a pure initial state, we pass the input state
$\ket{\phi}$ through a quickly spinning HWP preceded and followed
by two quarter-waveplates (QWP), which produces the state
\be
{\rho}_{\mathcal{S}}  = 
\begin{pmatrix}
\cos^2{\phi} && C \Omega/\omega \\
C^* \Omega/\omega && \sin^2{\phi}
\end{pmatrix},
\ee

where $|C| \leq \sin{\phi}\cos{\phi}$, $\omega$ is angular velocity of the spinning waveplate and $\Omega$ is the camera integration time. When the condition of $\omega \gg \Omega$ is satisfied, it results in a mixed state given by ${\rho}_{\mathcal{S}}(\phi)=\sin^{2}{(\phi)}\ket{H}\bra{H}+\cos^{2}{(\phi)}\ket{V}\bra{V}$. This way, the intensity of the path modes depends on the angle $\phi$. 

The analysis of coherences is more accurate if intensities are non-vanishing, we further transform the state to
\be  \label{rotated}
{\rho}_{\mathcal{S}} = \frac{1}{2}
\begin{pmatrix}
1 && \cos{2 \phi} \\
\cos{2 \phi}  && 1
\end{pmatrix},
\ee
using a HWP at 22.5$^{\circ}$ (a rotation by $\pi/4$). After this transformation, the paths have equal and constant intensities, causing the state reconstruction process to be more accurate. The purity ${\rm Tr}(\rho^2)$ of the quantum states is unaffected by this transformation.

\section{SPDC}
As our single-photon source, we used a 15 mm periodically poled Potassium Titanyl Phosphate crystal for a Spontaneous Parametric Down-Conversion (SPDC) source, pumped by a diode laser at $\lambda= 404$ nm and producing photons pairs at $\lambda= 808$ nm. The measured $g^{2}(0)$ for the source was $g^{2}(0)=0.1979 \pm 0.0005$. This is the same source used in Ref.~\cite{Curic19}, where the high-dimension experimental tomography of a path-encoded photon quantum state was first introduced.

\section{Quantum state reconstruction}

Techniques for reconstructing the density matrix of a physical system (quantum state tomography) depend on the physical substrate used~\cite{DArianoetal2003,PRA_Measurement_of_qubits,Rippeetal2008,Lundeenetal2009}. Here we describe an approach that reconstructs the density matrix of a quantum state of a single photon that is in a superposition of multiple beam paths \cite{Curic19, Milburn1989,Recketal1994,Cerfetal1998}. In the present work, we adapted the method of Curic \textit{et al}. \cite{Curic19} as described below.

To obtain the diagonal elements of the density matrix, we either retrieve the diagonals from non-interfering paths after the 1D Optical Fourier transform (OFT), or we divide the intensity of each path by the total intensity. Because the horizontal direction has repeated spacings (the left-most two and right-most two have a spacing of 2.7 mm as shown in Fig.~\ref{fig:conceptual}), the Fourier method described in Ref.~\cite{Curic19} must be modified. While the coherences between vertically separated paths $\{a_{18}, a_{27}, a_{36},a_{45}\}$ can easily be recovered by aligning the OFT-axis at 90$^{\circ}$ with respect to the horizontal, coherences between horizontally separated path modes (such as $a_{87}$ and $a_{56}$) cannot be resolved with this method due to their equal spacing. One way to fix this issue is to block certain paths. As such, blocking the two left-most paths $a_{11}$ and $a_{88}$ resolves the issue of equal spacing. 

We can also exploit the fact that the coherences between elements along the horizontal are expected to be zero both in the collapse and the no-collapse picture. If we block $a_{11}$ and $a_{88}$, the remaining beam paths along the horizontal have a different spacing. In cases where the theory predicts zero coherence, it is still possible that a small residual non-zero value be detected due to experimental imperfections and detection noise. Experimental errors can be attributed to non-ideal beamsplitters and imperfections on the calcite crystals.

The reconstruction of the states starts with a 4$f$-system ($f_1 = 1000$ mm and $f_2 = 400$ mm) imaging the 8 path modes obtained after the third calcite into the the Electron-Multiplying CCD (EMCCD) camera. Optical Fourier transform (OFT) along the OFT-axis is performed by rotating a 250~mm cylindrical lens, and the coherence value is obtained via Discrete Fourier Transform (DFT) of the interference patterns.

 \section{Number of experimental runs}
 While we planned for five experimental replicates for all $\phi$ values, due to a problem in the data acquisition some angles ended up with fewer experimental runs. We report the number of replicates for each value of $\phi$ in Table~\ref{number_exprimental_runs}, for both the walk-off compensated (that is, coherent) as well as uncompensated (incoherent) experiments.

\begin{table}[h]
    \centering
    \begin{tabular}{|c|c|c|}
    \hline
     $\phi$ & Replicates (compensated)&  Replicates (non-compensated)\\
    \hline
    $0^\circ$ &4  & 5  \\
    \hline
    $4.5^\circ$& 5  & 5 \\
    \hline
    $9^\circ$  & 5 & 5 \\
    \hline
    $13.5^\circ$& 5  & 5 \\
    \hline
    $18^\circ$ & 5  & 5 \\
    \hline
    $22.5^\circ$ & 5  & 5 \\
    \hline
    $27^\circ$ &5  & 5 \\
    \hline
    $31.5^\circ$& 5  & 5 \\
    \hline
    $36^\circ$ & 5  & 5 \\
    \hline
    $40.5^\circ$& 5  & 5 \\
    \hline
    $45^\circ$ & 2 & 5 \\
    \hline
    \end{tabular}
    \caption{Number of replicate experiments for each value of $\phi$, with walk-off compensation to maintain coherence, and without.}
    \label{number_exprimental_runs}
\end{table}

\section{Derivation of $\Tilde{\rho}_{123}$ and purity given a rotated initial state} \label{Approt}

To experimentally see the visibility of the interference fringes and to be away from zero intensity (which in practice is limited to the detection noise), we rotate the initial state by $\theta=\frac\pi4$, that is, the basis states were not the standard $H$ and $V$ states, but rather the diagonal states $D$ and $\bar D$. Here we demonstrate that while the final density matrix $\Tilde{\rho}_{123}$ (Eq.~\ref{rho_123_tilde}) is different than $\rho_{123}$ (Eq.~\ref{joint}), the purity remains the same, that is, Eq.~(\ref{purity}) is the same as Eq.~(\ref{purity_rho_123_tilde}). 

We begin with the initial quantum state in the $H,V$ basis
\be
\rho_\phi=\sin^2(\phi)|H\ra\la H|+\cos^2(\phi)|V\ra\la V|\;. \label{drho1}
\ee
Instead of measuring in this basis (which gives rise to the density matrix of the first measurement, Eq.~(1)), we will measure in the diagonal basis
\be
|D\ra=\frac1{\sqrt2}(|H\ra+|V\ra)\;\;,\;\; |\bar D\ra=\frac1{\sqrt2}(|H\ra-|V\ra)\;.
\ee
To prepare for this, first rewrite (\ref{drho1}) in the diagonal basis 
\be
\rho_\phi&=&\sin^2(\phi)\frac12(|D\ra+|\bar D\ra)(\la D| +\la \bar D|)\\
&&+\cos^2(\phi)\frac12(|D\ra-|\bar D\ra)(\la D| -\la \bar D|)\\
&=&\frac12   \begin{pmatrix}
      1 & \cos(2\phi) \\
      \cos(2\phi) & 1
   \end{pmatrix}\;.
\ee
In order to perform the first measurement, we have to purify the density matrix using the reference state $R$. Introducing the orthogonal reference state basis states $|d\ra=\sq(|h\ra+|v\ra)$ and $|\bar d\ra=\sq(|h\ra-|v\ra)$, we can write
\be
|QR\ra&=&\frac{\cof}{\sqrt2}(|D\ra|d\ra+|\bar D\ra|\bar d\ra)+\frac{\sif}{\sqrt2}(|D\ra|d\ra-|\bar D\ra|\bar d\ra)\\
&=&\sq\biggl(|D\ra\bigl[\cof|h\ra+\sif |v\ra\bigr]+|\bar D\ra\bigl[\cof|h\ra-\sif |v\ra\bigr]\biggr)\;.
\ee
We now measure in the diagonal basis by projecting onto $D$ and $\bar D$, using the ancilla states $|0\ra_1$ and $|1\ra_1$:
\be
|\psi\ra_1=\sq\biggl(|D\ra|0\ra_1(\cof|h\ra+\sif |v\ra)+|\bar D\ra|1\ra_1(\cof|h\ra-\sif |v\ra)\biggr)\;.
\ee
The density matrix of the first measurement device is
\be
\rho_1&=&\tr_{QR}|\psi_1\ra\la\psi_1|=\frac12|0\ra_1\la0|(\cofs+\sifs)+|1\ra_1\la1|(\cofs+\sifs)\nonumber \\
&=&   \begin{pmatrix} 
      \frac12 & 0 \\
      0 & \frac12 \\
   \end{pmatrix}\;.
\ee
The second measurement will be in a basis rotated by $45^\circ$, that is, in the $H,V$ basis. To do this, first rewrite the $D,\bar D$ basis states in the $H,V$ basis:
\be
|\psi_1\ra&=&\frac12\biggl(|H\ra\bigl[\cof|h\ra(|0\ra_1+|1\ra_1)+\sif|v\ra(|0\ra_1-|1\ra_1)\bigr]\\
&&\ \ \ \ \ +|V\ra\bigl[\cof|h\ra(|0\ra_1-|1\ra_1)+\sif|v\ra(|0\ra_1+|1\ra_1)\bigr]\biggr)\;.
\ee
Now measure $|\psi_1\ra$ using the ancilla for the second measurement $|0\ra_2$ and $|1\ra_2$. This produces
\be
|\psi_2\ra&=&\frac12\biggl(|H\ra|0\ra_2\bigl[\cof|h\ra(|0\ra_1+|1\ra_1)+\sif|v\ra(|0\ra_1-|1\ra_1)\bigr]\nonumber \\
&&\ \ \ \ \ +|V\ra|1\ra_2\bigl[\cof|h\ra(|0\ra_1-|1\ra_1)+\sif|v\ra(|0\ra_1+|1\ra_1)\bigr]\biggr)\;.
\ee
Tracing out the quantum state and the reference yields the joint density matrix for the first two measurements:
\be
\rho_{12}=\frac14   \begin{pmatrix} 
      1 & \cos(2\phi)& 0& 0 \\
       \cos(2\phi)& 1 & 0 &0 \\
       0& 0& 1& -\cos(2\phi)\\
       0 & 0& -\cos(2\phi) & 1
   \end{pmatrix}\;.
   \ee
Note that it shows non-zero off-diagonal elements, unlike in the case where the first measurement was carried out in the $H,V$ basis.

We can check the purity of this matrix. Since
\be
\rho_{12}^2=
 \frac1{16}\begin{pmatrix}
      1+\cos^2(2\phi) & 2\cos(2\phi)& 0& 0 \\
       2\cos(2\phi)& 1+\cos^2(2\phi) & 0 &0 \\
       0& 0& 1+\cos^2(2\phi)& -2\cos(2\phi)\\
       0 & 0& -2\cos(2\phi) & 1+\cos^2(2\phi)
   \end{pmatrix}\;.
   \ee
we obtain
\be
C_2=\tr\rho_{12}^2=\frac14(1+\cos^2(2\phi))\;,
\ee
which is identical to the purity of the two-measurement density matrix in the original $(H,V)$ basis.

We now move to the third and last measurement. As the third measurement is performed at an angle of $45^\circ$ to the second, we need to measure in the diagonal basis again. To do this, rewrite $|\psi\ra_2$ in that basis first:
\be
|\psi\ra_2&=&\frac12\sq\biggl((|D\ra+\bar D\ra)\ra|0\ra_2\bigl[\cof|h\ra(|0\ra_1+|1\ra_1)+\sif|v\ra(|0\ra_1-|1\ra_1)\bigr]\nonumber \\
&&\ \ \ \ \ +(|D\ra-\bar D\ra)\ra|1\ra_2\bigl[\cof|h\ra(|0\ra_1-|1\ra_1)+\sif|v\ra(|0\ra_1+|1\ra_1)\bigr]\biggr)\;.
\ee
Measuring in the $D,\bar D$ basis and tracing out the reference gives the full density matrix (with the abbreviation $c=\cos(2\phi)$):
\be
\rho_{123}=\frac18   \begin{pmatrix} 
      1 & 1 & c & -c & 0 & 0 & 0 & 0 \\
 1 & 1 & c & -c & 0 & 0 & 0 & 0 \\
 c & c & 1 & -1 & 0 & 0 & 0 & 0 \\
 -c & -c & -1 & 1 & 0 & 0 & 0 & 0 \\
 0 & 0 & 0 & 0 & 1 & -1 & c & c \\
 0 & 0 & 0 & 0 & -1 & 1 & -c & -c \\
 0 & 0 & 0 & 0 & c & -c & 1 & 1 \\
 0 & 0 & 0 & 0 & c & -c & 1 & 1 \\
   \end{pmatrix}\;.
\ee
This is the matrix used to compare experimental results to in Figs. 5 and 7-10.
The trace of that matrix is unity, and its purity is
\be
C_3=\tr(\rho_{123}^2)=\frac14(1+\cos^2(2\phi))\;,
\ee
precisely as when making the measurements on the diagonal initial state in the $H,V$ basis (compare to Eq.~(\ref{purity})).

\section{Reconstructed density matrices after three measurements for different values of $\phi$}
\label{extramat}
\begin{figure}[]   
   \centering
   \includegraphics[width=1.0 \textwidth]{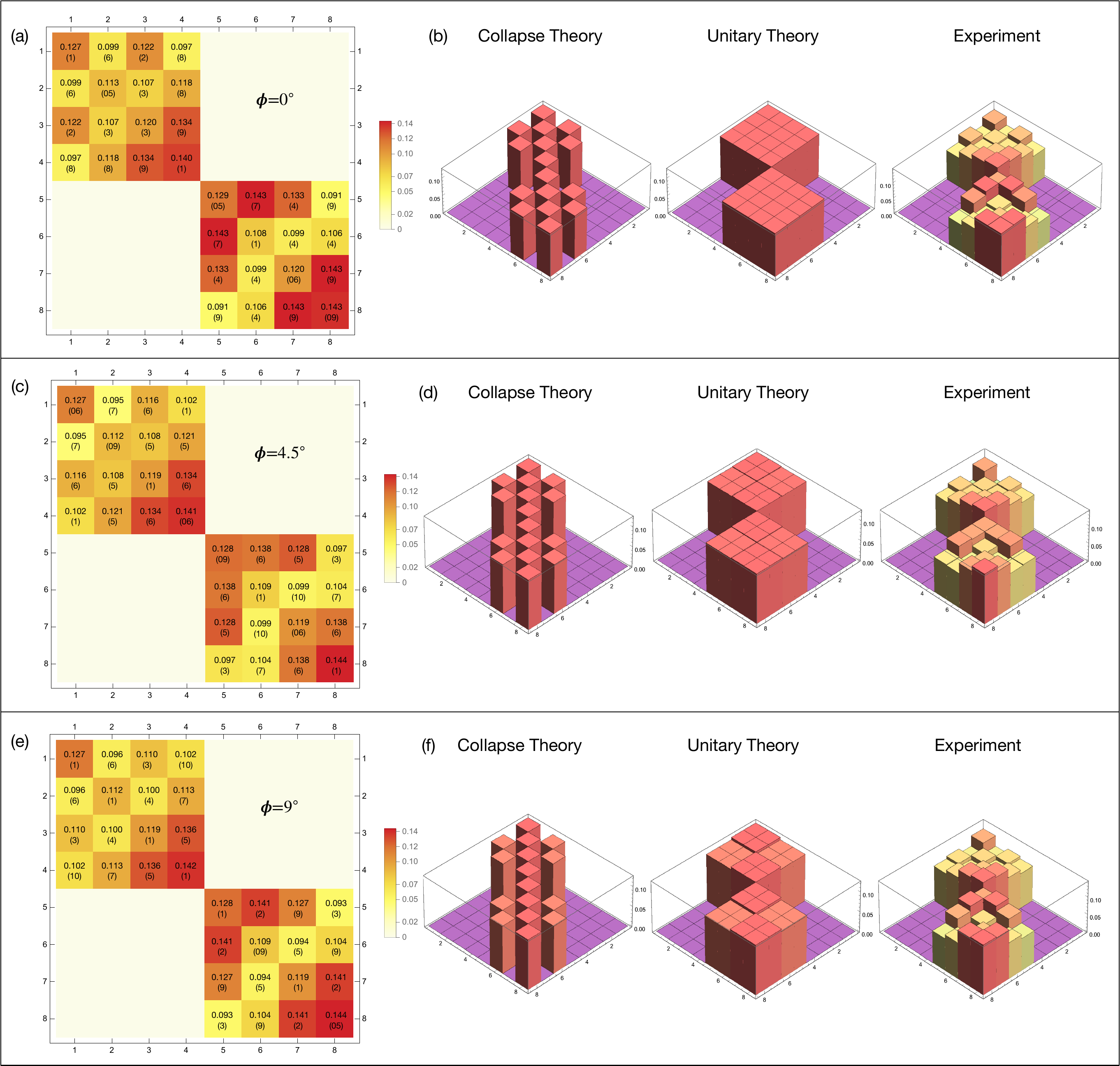} 
   \caption{Results of the reconstruction of the density matrix using after three consecutive measurements. 
   (a) Experimental value of the corresponding element of the joint density matrix after three consecutive measurements [Eq.~(\ref{rho_123_tilde})], averaged over experimental replicates (see Appendix D, Table II), with one standard deviation expressed in concise form, for $\phi=0^\circ$. (b) Prediction of the density matrix values rendered as a Manhattan plot of the collapse theory, the unitary theory, and the experimental values (the latter are the same as shown numerically in (a)).
   (c) Experimental results for $\phi=4.5^\circ$, (d) corresponding Manhattan plots at $\phi=4.5^\circ$.
   (e)  Experimental results for $\phi=9^\circ$, (f) corresponding Manhattan plots at $\phi=9^\circ$.
   }
   \label{extra1}
\end{figure}

\begin{figure}[]   
   \centering
   \includegraphics[width=1.0 \textwidth]{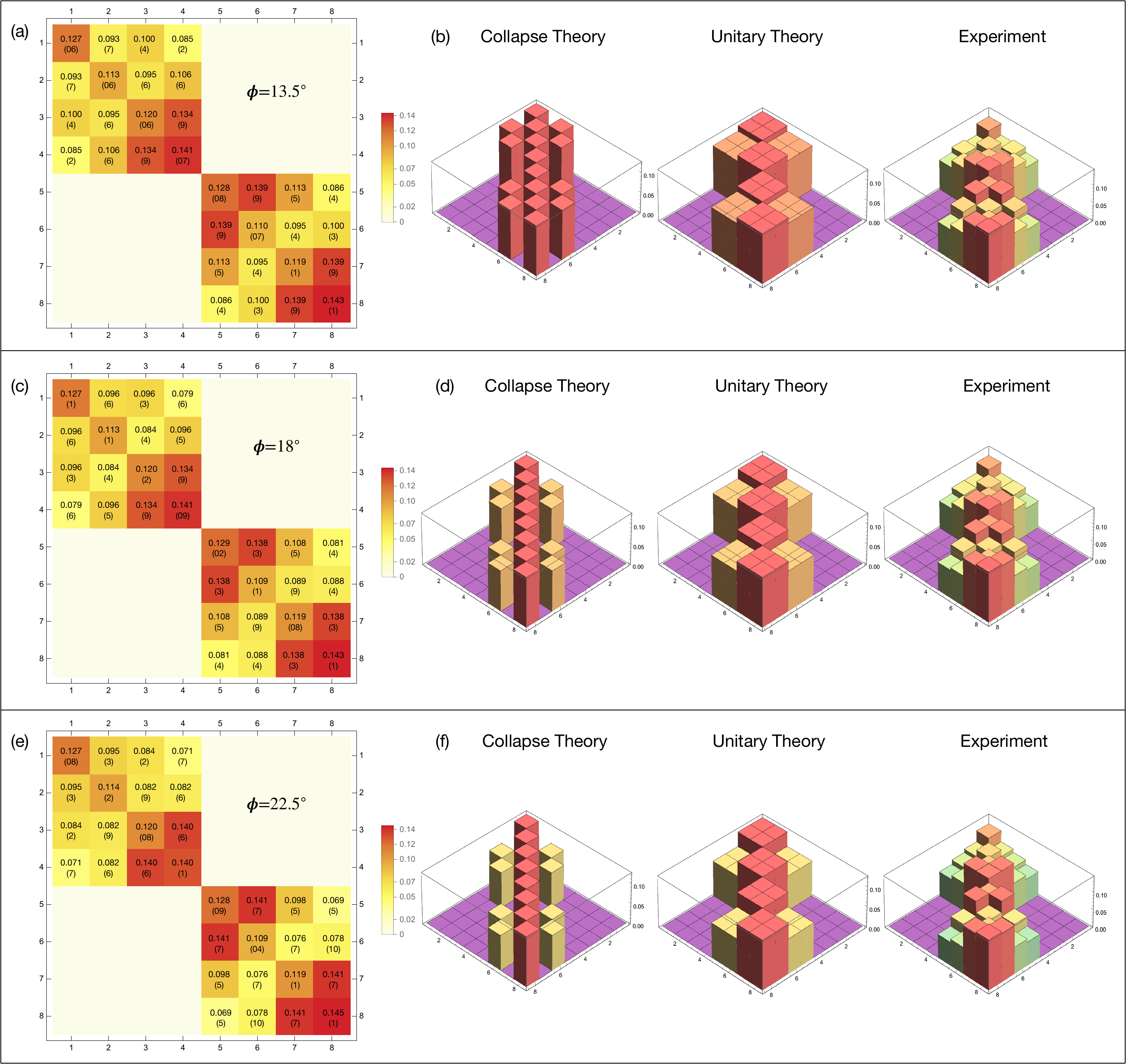} 
   \caption{Results of the reconstruction of the density matrix after three consecutive measurements. 
   (a) Experimental value of the corresponding element of the joint density matrix after three consecutive measurements [Eq.~(\ref{rho_123_tilde})], averaged over experimental replicates (see Appendix D, Table II), with one standard deviation expressed in concise form, for $\phi=13.5^\circ$. (b) Prediction of the density matrix values rendered as a Manhattan plot of the collapse theory, the unitary theory, and the experimental values (the latter are the same as shown numerically in (a).
   (c) Experimental results for $\phi=18^\circ$, (d) corresponding Manhattan plots for $\phi=18^\circ$.
   (e)  Experimental results for $\phi=22.5^\circ$, (f) corresponding Manhattan plots for $\phi=22.5^\circ$.}
\end{figure}

\begin{figure}[]   
   \centering
   \includegraphics[width=1.0 \textwidth]{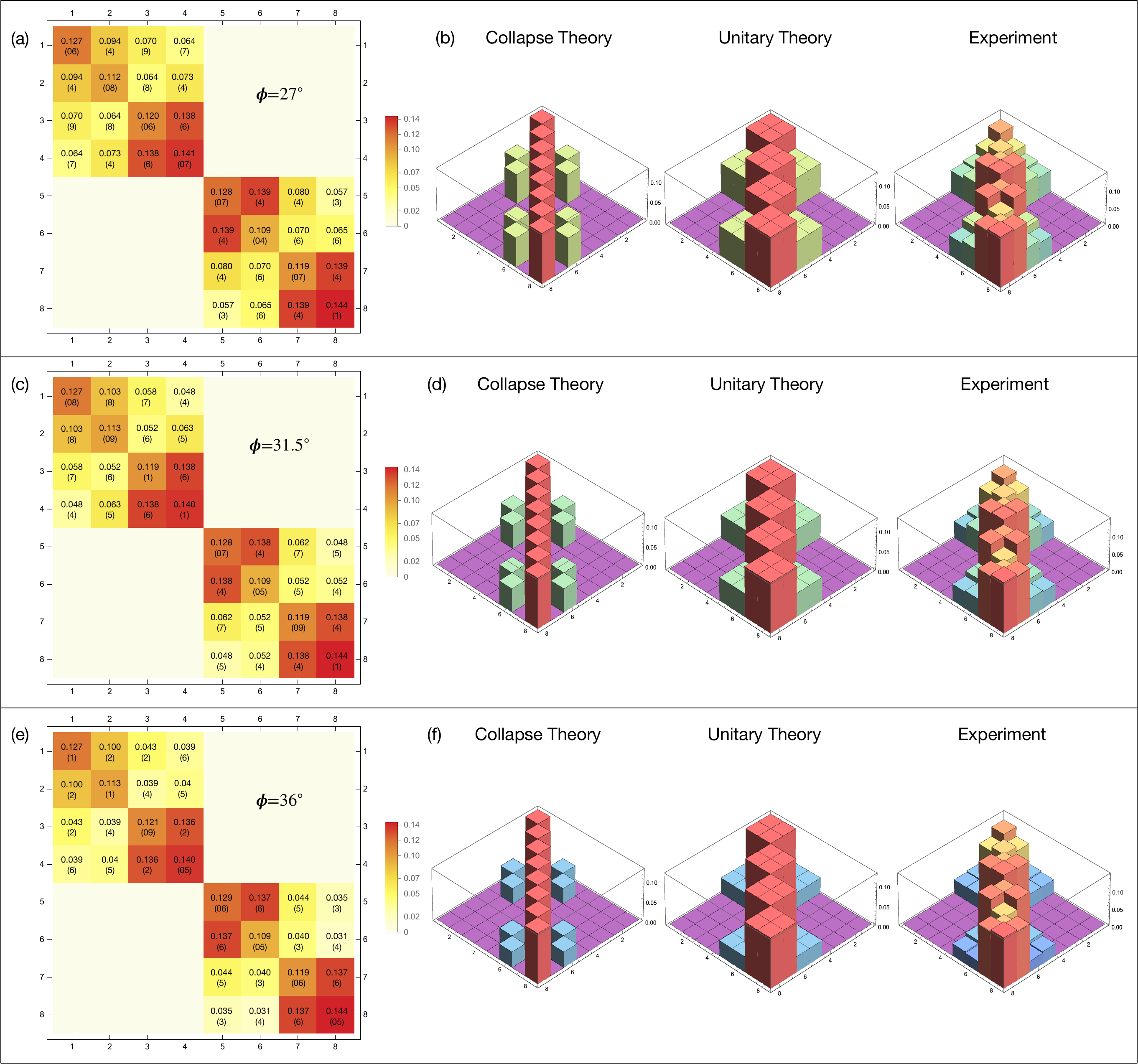} 
   \caption{Results of the reconstruction of the density matrix after three consecutive measurements. 
   (a) Experimental value of the corresponding element of the joint density matrix after three consecutive measurements [Eq.~(\ref{rho_123_tilde})], averaged over experimental replicates (see Appendix D, Table II), with one standard deviation expressed in concise form, for $\phi=27^\circ$. (b) Prediction of the density matrix values rendered as a Manhattan plot of the collapse theory, the unitary theory, and the experimental values (the latter are the same as shown numerically in (a).
   (c) Experimental results for $\phi=31.5^\circ$, (d) corresponding Manhattan plots for $\phi=31.5^\circ$.
   (e)  Experimental results for $\phi=36^\circ$, (f) corresponding Manhattan plots for $\phi=36^\circ$.}
\end{figure}

\begin{figure}[]   
   \centering
   \includegraphics[width=1.0 \textwidth]{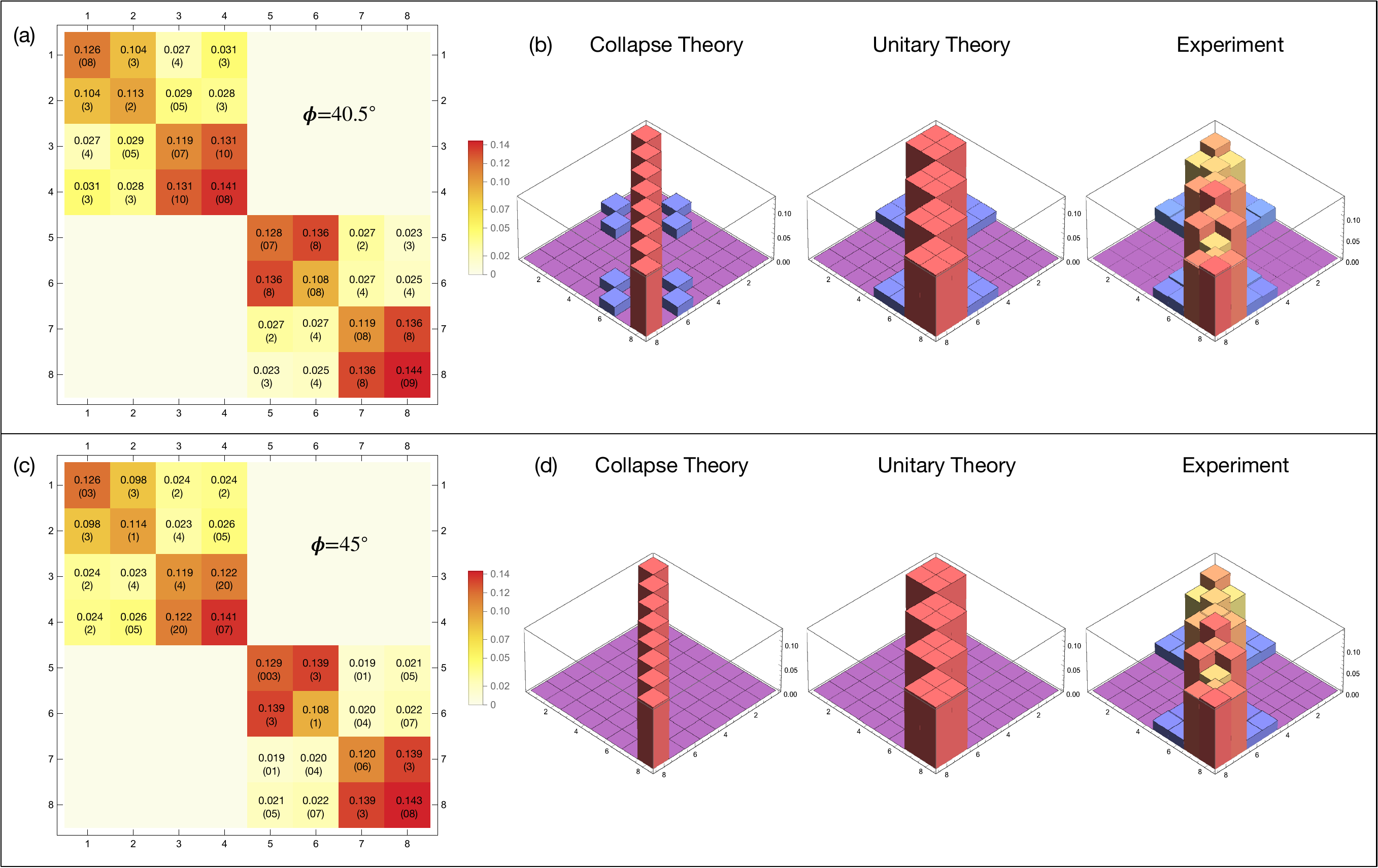} 
   \caption{Results of the reconstruction of the density matrix after three consecutive measurements. 
   (a) Experimental value of the corresponding element of the joint density matrix after three consecutive measurements [Eq.~(\ref{rho_123_tilde})], averaged over experimental replicates (see Appendix D, Table II), with one standard deviation expressed in concise form, for $\phi=40.5^\circ$. (b) Prediction of the density matrix values rendered as a Manhattan of the collapse theory, the unitary theory, and the experimental values (the latter are the same as shown numerically in (a).
   (c) Experimental results for $\phi=45^\circ$, (d) corresponding Manhattan plots for $\phi=45^\circ$.}
\end{figure}
\end{widetext}

\clearpage
\bibliography{quant}
\bibliographystyle{unsrt}
\bibliographystyle{ieeetr}

\end{document}